\documentclass[12pt]{iopart}

\usepackage{graphicx}

\begin{document}

\title[Cosmic Ray Propagation in the Interstellar Magnetic Fields]{Cosmic Ray Propagation in the Interstellar Magnetic Fields}

\author{Gwenael Giacinti}

\address{University of Oxford, Clarendon Laboratory, Parks Road, Oxford OX1 3PU, UK}

\ead{gwenael.giacinti@physics.ox.ac.uk}
\begin{abstract}
The propagation of TeV--PeV cosmic rays (CR) in our Galaxy can be described as a diffusive process. We discuss here two effects, with important observational consequences, that cannot be predicted by the diffusion approximation (DA) in its usual form. First, we present an explanation for the CR anisotropies observed at small angular scales on the sky. We show that the local magnetic field configuration within a CR mean free path from Earth naturally results in CR flux anisotropies at small and medium scales~\cite{Giacinti:2011mz}. Second, we point out that TeV--PeV CRs should be expected to diffuse strongly anisotropically in the interstellar medium on scales smaller than the maximum scale of spatial fluctuations of the field, $\sim100$\,pc~\cite{Giacinti:2012ar,Giacinti:2013wbp}. This notably questions the usual assumptions on CR diffusion around sources.
\end{abstract}

%\maketitle

\section{Introduction}

The turbulent magnetic field that permeates the interstellar medium (ISM) of our Galaxy contains fluctuations on spatial scales ranging from $L_{\max} \sim 100$\,pc down to a damping scale $L_{\min}$, smaller than $\sim 1$\,AU~\cite{Han:2004aa}. Galactic cosmic rays (CR) resonantly scatter on modes with fluctuation scales that match their Larmor radius $r_{\rm L}$ : $2\pi/k \sim r_{\rm L}$. In turn, individual trajectories of CRs may be modeled as random walks. CR propagation in the disk and halo can be globally described as a diffusive process, by the diffusion approximation (DA) ---see for instance Refs.~\cite{Ginzburg:1990sk,galprop} for more exhaustive discussions.

Streaming CRs generate Alfv\'en waves~\cite{Skilling1,Skilling2}. However, above energies of a few tens of GeVs, CRs should not be expected to be self-confined, and, hence, do not stream in the ISM to only $\sim$ the Alfv\'en speed~\cite{Skilling2,Wentzel}. The exact impact of higher energy CRs on the magnetic fields they diffuse in is still an open question. The authors of Refs.~\cite{Blasi:2012yr} argue that the spectral inflection detected by PAMELA and CREAM experiments~\cite{PAMELACREAM} at rigidities $E/Z \sim 200$\,GeV could mark the limit between CR diffusion in self-generated turbulence and diffusion in the pre-existing turbulence. In Galactic CR propagation codes~\cite{galprop,Evoli:2008dv}, this can be implemented by assuming a break in the diffusion coefficient at that rigidity, but it does not change the global effective picture of CR diffusion in the ISM. We discuss in this work two topical subjects that cannot be predicted by the DA in its usual form.

Several observatories have detected anisotropies in the TeV--PeV CR arrival directions on the sky at Earth~\cite{CRAobs1,CRAobs2}. These anisotropies have been observed both at large (dipole) and, more surprisingly, small angular scales (higher order multipoles). The amplitude of the dipole is $\simeq 0.1$\%, and that of the smaller scale anisotropies (SSA) is lower by a factor of a few to ten. SSA have been detected down to $\simeq10^\circ$ scales, with narrow regions receiving a flux larger or smaller than would be expected if only a dipole anisotropy were present. The DA cannot explain the SSA. We show in Section~\ref{CRASAS} that it is so because the DA is not designed to predict phenomena arising on spatial scales smaller than the CR mean free path (MFP) : Anisotropies naturally appear on small angular scales and reflect the local configuration of the turbulent magnetic field within a CR MFP from Earth.

In Section~\ref{ACRD}, we study the diffusion of high-energy Galactic CRs around their sources. We show that the simplifying hypothesis of isotropic diffusion is usually not acceptable. CR densities around sources should be expected to be irregular and, sometimes, to display filamentary structures.

\section{Origin of TeV--PeV cosmic ray anisotropies}
\label{CRASAS}

The DA only predicts a dipole anisotropy. The direction and amplitude of the dipole $\hbox{\boldmath$\delta$}$ are directly related to the local relative gradient of CR density $\hbox{\boldmath$\nabla$}n/n$ by $\hbox{\boldmath$\delta$}=3D(E)/c \cdot \hbox{\boldmath$\nabla$}n/n$, where $D(E)$ is the CR diffusion coefficient. $\hbox{\boldmath$\nabla$}n/n$ is mostly determined by one or a few nearby recent sources~\cite{CRADipole}. Refs.~\cite{DE}, amongst others, put constraints on the amplitude of $D(E)$. Taking values of $D$ in the preferred range, and $|\hbox{\boldmath$\nabla$}n/n| \sim 1$\,kpc$^{-1}$, one finds $|\hbox{\boldmath$\delta$}| \sim 0.1$\%~\cite{Giacinti:2011mz}, which is compatible with observations. See Refs.~\cite{CRADipole} for more detailed calculations.

However, the DA does not predict the SSA. Previous works have assumed additional effects so as to explain the SSA, see for example Refs.~\cite{SSAth}. We demonstrate here that no additional effect is needed, and that rigidity-dependent SSA automatically arise from the local realization of the random magnetic field within about a CR MFP from Earth. This statement is valid even in isotropic magnetic turbulence. In other words, the DA does not predict SSA because it loses its validity on distances smaller than the CR MFP.

Let us schematically model CR trajectories as random walks. CRs can then be regarded as performing jumps on straight lines with MFP lengths, $l_{\rm mfp}$. They have an equal probability to go in any direction after every jump because the process is Markovian. In this simplified picture, CRs arriving at Earth during a time interval $l_{\rm mfp}/c$ are located within a sphere of radius $l_{\rm mfp}$. Assuming a gradient of CR density $\hbox{\boldmath$\nabla$}n$ at Earth, one then only finds a dipole anisotropy in the direction of $\hbox{\boldmath$\nabla$}n$, with the amplitude predicted by the DA. However, in reality, the magnetic turbulence has a given configuration, and CRs do not travel on straight lines within this sphere of radius $\approx l_{\rm mfp}$. The magnetic field within $\approx l_{\rm mfp}$ from Earth does not vary significantly during the CR crossing time $\sim l_{\rm mfp}/c$. This gives rise to anisotropies at medium and small scales by reshuffling pieces of the dipole to smaller scales, on different parts of the sky. Equivalently, one can say that the DA cannot predict the SSA because it averages over all possible magnetic field configurations. This averaging notably causes the problem to be artificially symmetric around the axis defined by $\hbox{\boldmath$\nabla$}n$ and containing the Earth.

\begin{figure}
\begin{center}
\includegraphics[width=0.49\textwidth]{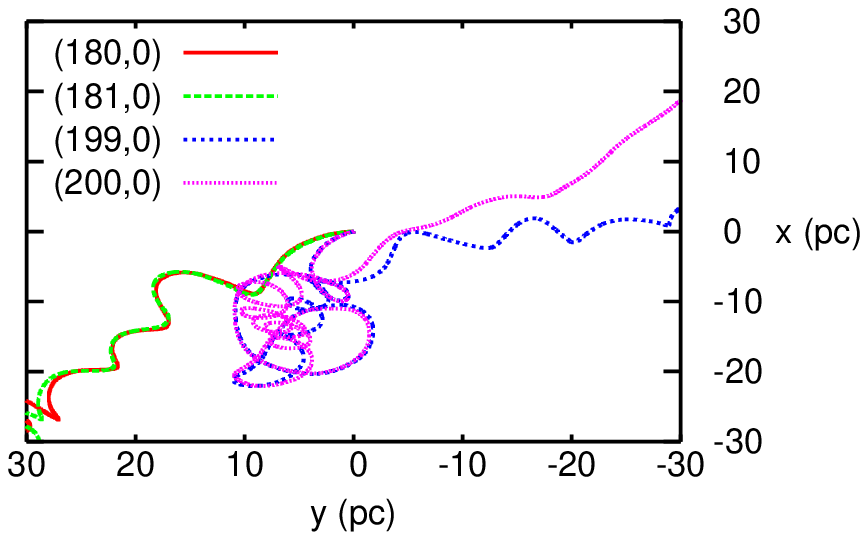}
\includegraphics[width=0.49\textwidth]{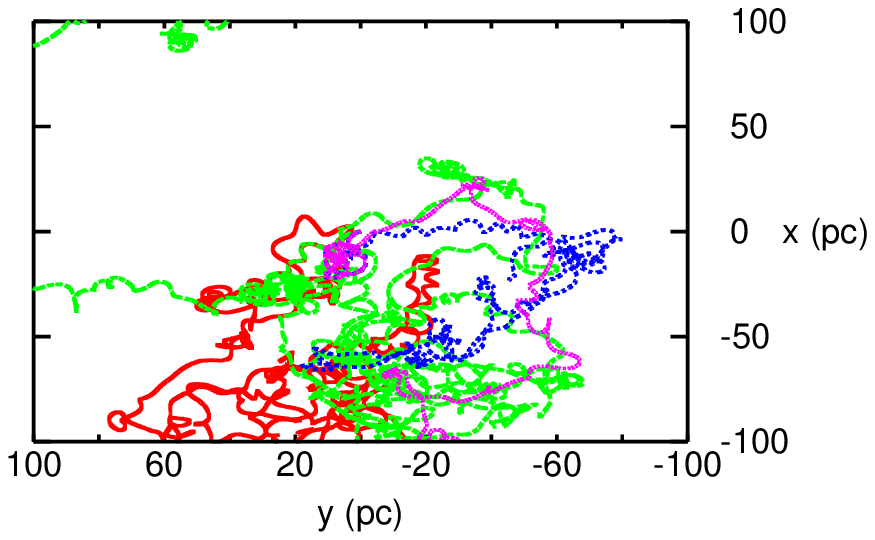}
\end{center}
\caption{Simulated trajectories of four CRs, with rigidities $E/Z=10$\,PeV, arriving at Earth in different directions on the sky: ($l,b$) = (180$^{\circ}$,0$^{\circ}$), (181$^{\circ}$,0$^{\circ}$), (199$^{\circ}$,0$^{\circ}$) and (200$^{\circ}$,0$^{\circ}$) --See key. Trajectories are projected on the Galactic plane (x,y) in regions with 60\,pc$\times$60\,pc {\it (left panel)} and 200\,pc$\times$200\,pc {\it (right panel)} sizes. The Earth is located at (0,0) and $\hbox{\boldmath$\nabla$}n$ points towards increasing y. Concrete realization of the turbulent magnetic field, see text for its parameters.}
\label{Traj}
\end{figure}

Figure~\ref{Traj} presents the trajectories of four CRs with $E/Z=10$\,PeV, propagated in isotropic magnetic turbulence. The magnetic field is precomputed on a 3D grid in physical space from its power spectrum in reciprocal space, using a Fast Fourier Transform. It is then interpolated in any point of the CR trajectories. Each vertex of the grid in reciprocal space corresponds to a wave vector {\bf k}, see~\cite{TFGen}. Since fluctuations in the field should be resolved, at least, down to spatial scales $\approx r_{\rm L}/10$, we use the nested grid method proposed in~\cite{Giacinti:2011ww}. This allows us to have dynamic ranges $L_{\max}/L_{\min}$ significantly larger than the number of vertexes per edge of individual grids. We use here $L_{\min}=2 \cdot 10^{-4}$\,pc, $L_{\max}=150$\,pc, and a Kolmogorov power spectrum $\mathcal{P}(k)\propto k^{-5/3}$ ---the amplitude of the Fourier modes follows $|{\bf B}({\bf k})|^{2} \propto k^{-11/3}$. The root mean square strength of the field is $B_{\rm rms}=4\,\mu$G, and no regular field is added. The four CRs in Fig.~\ref{Traj} hit the Earth, located at (0,0), and are chosen such that after subtracting the dipole, two of the trajectories ---with coordinates on the sky ($l,b$) = (180$^{\circ}$,0$^{\circ}$) and (181$^{\circ}$,0$^{\circ}$), are in a hot spot of the SSA and the two others in a cold spot. In the left panel, one can see that neighbouring trajectories stay close to one another within this region of size $\approx l_{\rm mfp}$, and that the two trajectories in the hot (resp. cold) spot go in the direction of higher (resp. lower) CR densities. This is in line with SSA being determined by the trajectories followed by CRs within $\approx l_{\rm mfp}$ from Earth~\cite{Giacinti:2011mz}. The right panel presents the same trajectories, in a larger region around Earth: As expected, trajectories that are close to one another in the left panel are not correlated here, on distances $>l_{\rm mfp}$ from Earth.

\begin{figure}
\begin{center}
\includegraphics[width=0.49\textwidth]{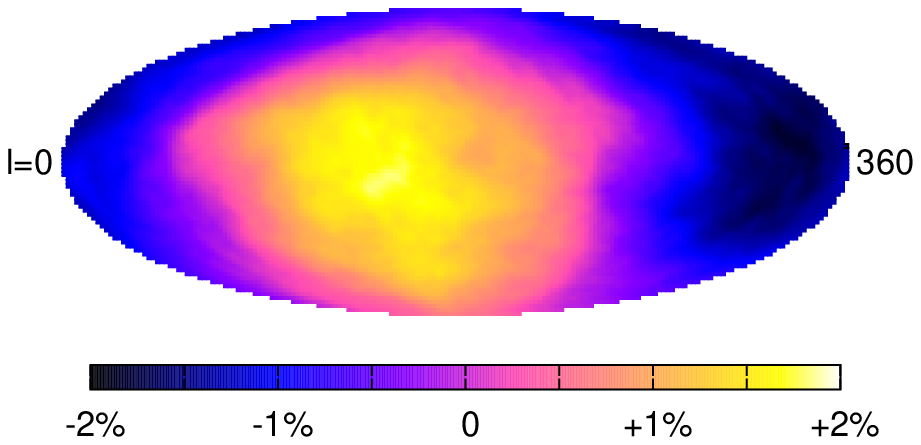}
\includegraphics[width=0.49\textwidth]{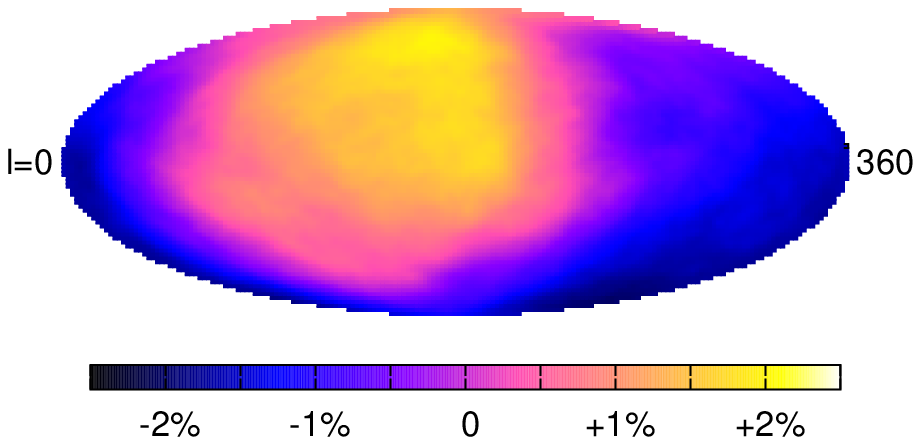}
\includegraphics[width=0.49\textwidth]{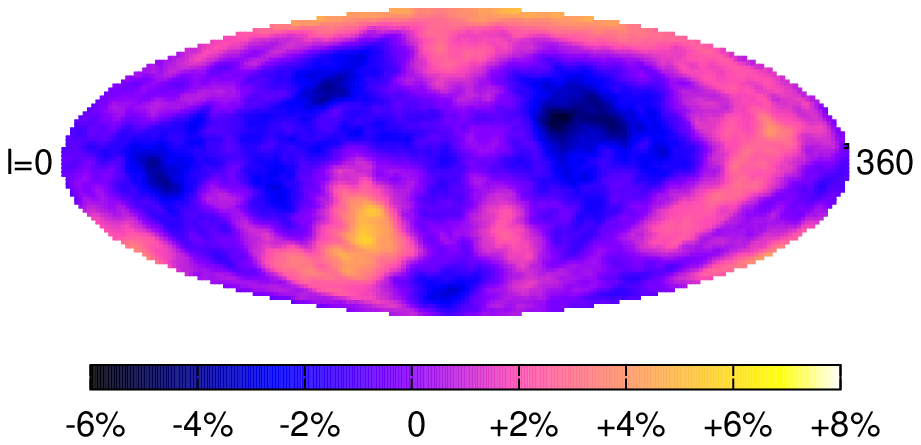}
\includegraphics[width=0.49\textwidth]{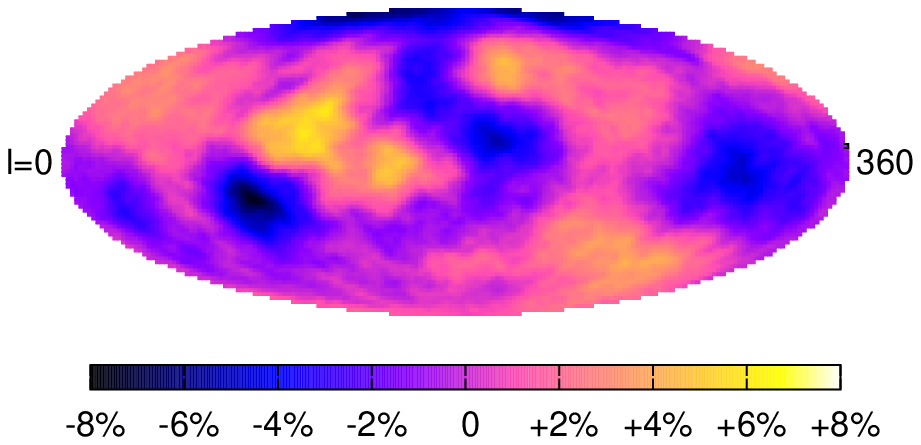}
\end{center}
\caption{Deviations from the average CR flux on the celestial sphere at Earth, when smoothing over 90$^{\circ}$ circles {\it before} {\it (upper row)}, and over 20$^{\circ}$ circles {\it after} subtracting the dipole {\it (lower row)}. The two columns present results for two different concrete realizations of the turbulent magnetic field. See text for values of parameters.}
\label{DipoleAndSmallerScales}
\end{figure}

Figure~\ref{DipoleAndSmallerScales} shows the CR flux smoothed over 90$^{\circ}$ circles on the sky (upper row) and the remaining SSA after subtracting the dipole and smoothing over 20$^{\circ}$ circles (lower row), for two different realizations of the magnetic turbulence (one for each column). For computing time reasons, we take CRs with $E/Z=10$\,PeV and a relative gradient of CR density $|\hbox{\boldmath$\nabla$}n/n|=(290\,{\rm pc})^{-1}$, which gives a dipole amplitude of $\simeq 6$\%. More realistic values would give an amplitude in line with those observed. The amplitude of SSA is about the same. In practice, CRs with different rigidities are mixed (different charges and broad energy ranges), which leads to a smaller amplitude because SSA are rigidity-dependent and values for different $E/Z$ add non-constructively. Taking a CR distribution with a median energy of 10\,PeV and a relative width $\Delta E/E$ equal to that inferred for IceCube measurements at 20\,TeV (see Fig.~3 of Ref.~\cite{CRAobs1}), we find SSA with an amplitude $\approx 10$ times lower than the dipole. This is in line with observations.

In Fig.~\ref{DipoleAndSmallerScales}, both dipoles should point towards (180$^{\circ}$,0$^{\circ}$), which is the direction of $\hbox{\boldmath$\nabla$}n$ here. Results are not far from the DA prediction, but slight deviations are visible, because of the local realization of the turbulence. Concerning the SSA, the two lower panels are completely different from one another. This illustrates the full dependence of SSA on the field realization.

\section{Anisotropic diffusion of cosmic rays in the ISM}
\label{ACRD}

\begin{figure}
\begin{center}
\includegraphics[width=0.49\textwidth]{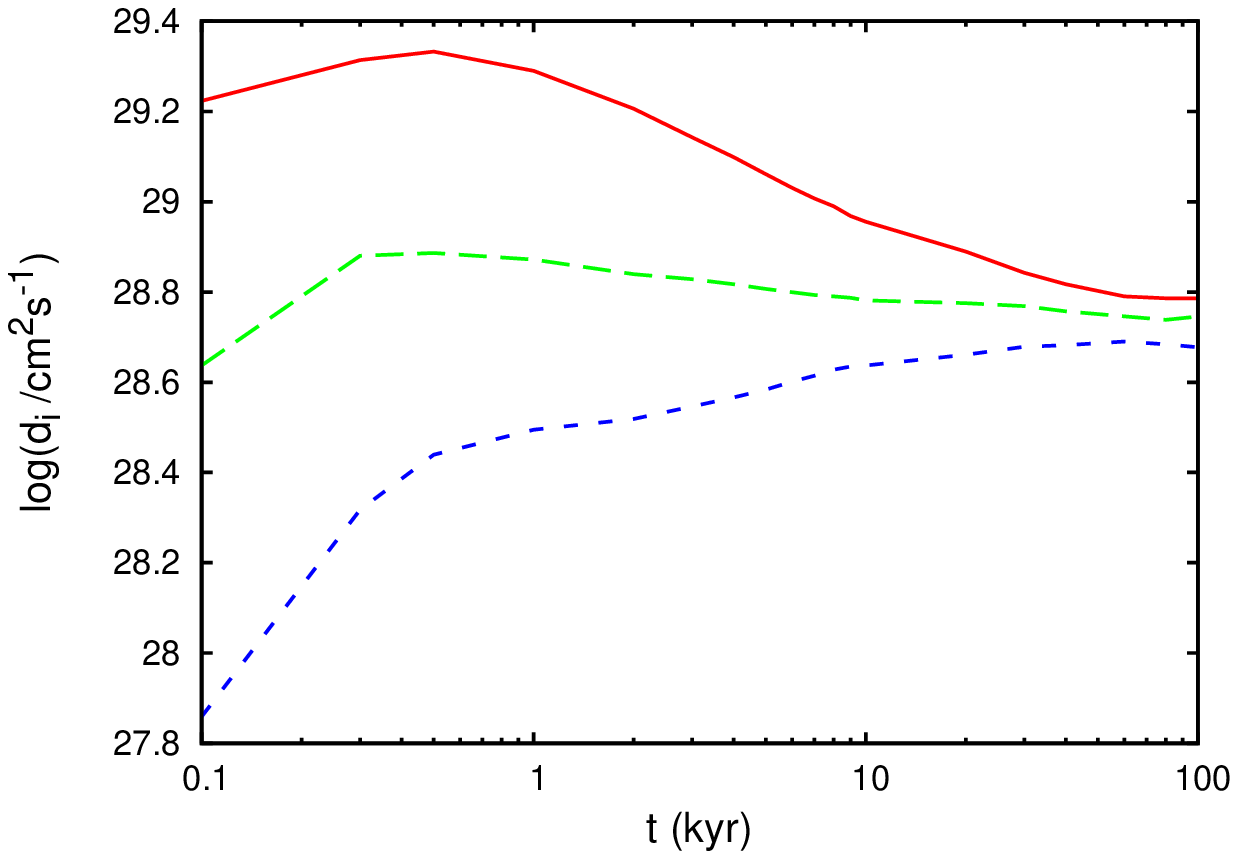}
\includegraphics[width=0.49\textwidth]{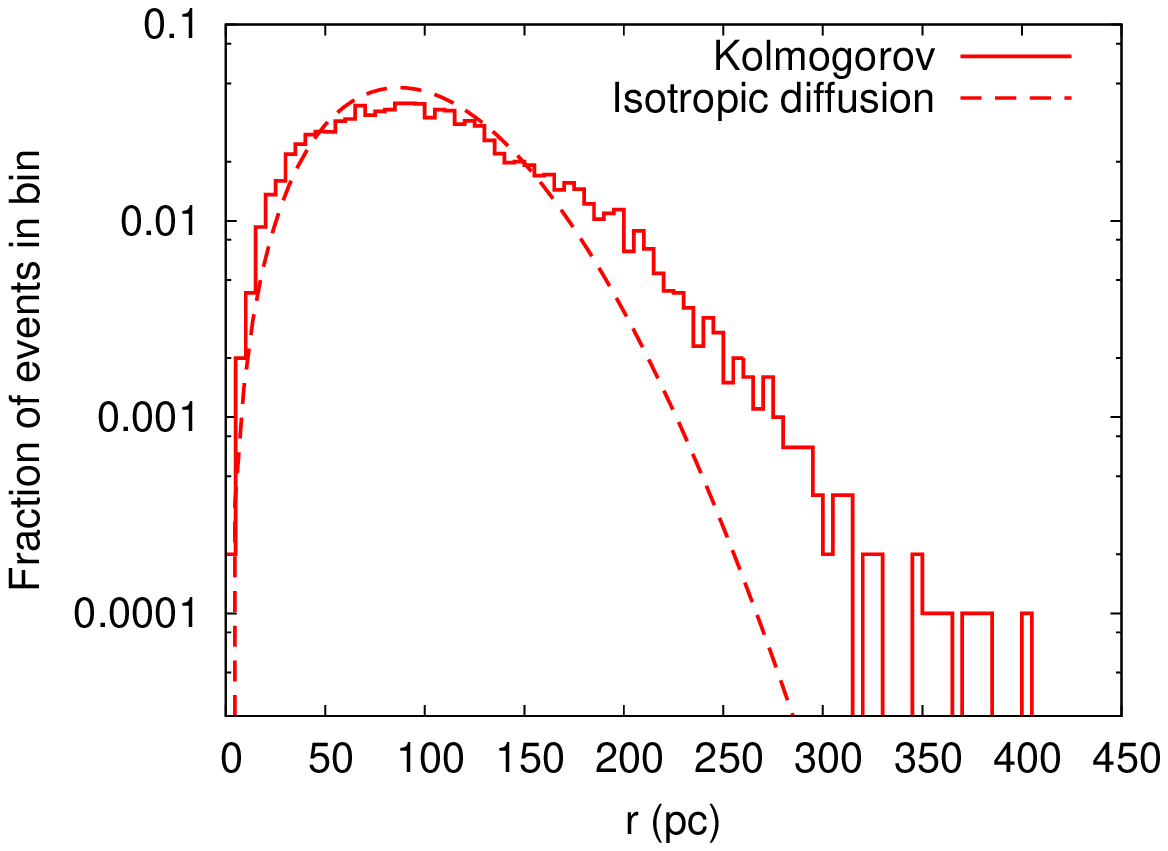}
\end{center}
\caption{{\it Left panel:} Average eigenvalues $d_{1,2,3}$ (see text for definition) versus time for CRs with $E/Z=1$\,PeV, emitted at (0,0,0); {\it Right panel:} Radial distribution of CRs with $E/Z=1$\,PeV at $t=t_\ast=10$\,kyr after emission at $r=0$ (solid line), compared with the expectation for isotropic diffusion (dashed line). CRs are propagated in turbulent field configurations with parameters identical to those of Fig.~\ref{Traj}.}
\label{Eigenval}
\end{figure}

Observations are consistent with either a Kolmogorov or a Kraichnan spectrum for the power spectrum of the turbulent Galactic magnetic field~\cite{galprop}. They suggest that the coherence length of the field, $L_{\rm c}$, is about a few tens of parsecs~\cite{Han:2004aa,Frisch:2010fw}. This implies that the Larmor radius of TeV--PeV CRs should be $\ll L_{\rm c},L_{\max}$\,, and that most of the power is contained in fluctuations on spatial scales larger than $r_{\rm L}$. Since the ratio of power in modes with $2\pi /k \sim r_{\rm L}$ to modes with $2\pi /k \gg r_{\rm L}$ is expected to be small, CRs should diffuse strongly anisotropically locally, because the latter modes are seen as local regular fields. As in Section~\ref{CRASAS}, we propagate individual CRs in turbulent fields. To quantify the anisotropy of CR distributions around sources, we compute the eigenvalues $d_{1}^{(b)}<d_{2}^{(b)}<d_{3}^{(b)}$ of $D_{ij}^{(b)}=\frac{1}{2nt} \sum_{a=1}^{n} x_i^{(a)}x_j^{(a)}$ for $n=10^4 \gg 1$ CRs with $E/Z=1$\,PeV, injected at (0,0,0) and $t=0$ in {\it one given} realization $b$ of the {\it isotropic} Kolmogorov turbulence described previously. The effects of variation over time of the field are negligible during the time scales we consider ($t \sim 10$\,kyr) : The velocity of ISM fluid parcels and the Alfv\'en speed are $\sim 10$\,km/s, and 10\,km/s\,$\times$\,10\,kyr$\,\ll \sqrt{d_{i}\times 10\,{\rm kyr}}$. The degree of anisotropy of CR distributions vary from one configuration $b$ to another. Therefore, we study the averages of $d_{1,2,3}^{(b)}$ over $N_b \gg 1$ different configurations of the field, $d_{k}=\frac{1}{N_b}\sum_{b=1}^{N_b}d_{k}^{(b)}$. Figure~\ref{Eigenval} (left panel) shows the evolution over time of $d_{1,2,3}$ averaged over $N_b = 10$ magnetic field realizations. $N_b = 10$ is sufficient for the purpose of this work. At early times, the ratio between extreme eigenvalues reaches about a few tens. Then, all eigenvalues tend towards the same value for $t \sim t_{\ast} \sim 10\,{\rm kyr}$. $t_{\ast}$ corresponds to the time when the bulk of spreading CRs reach a distance $\approx L_{\max}$ from the source, and therefore start to experience other 'cells' of size $L_{\max}^{3}$. Schematically, CRs are initially contained in a more or less narrow flux tube containing the source ($t \sim t_{\ast}/10$). The same effect can explain solar energetic particle dropouts in the solar wind, see~\cite{dropouts}. Then, CRs start to be more isotropized in space ($t \sim t_{\ast}$), though their radial distributions from their sources still differ at large radii $r$ from the predictions of isotropic diffusion, see Fig.~\ref{Eigenval} (right panel). The excess in the tail of the distribution at large $r$ becomes unnoticeable by $t \sim 10\,t_{\ast}$.

\begin{figure}
\begin{center}
\includegraphics[width=0.325\textwidth]{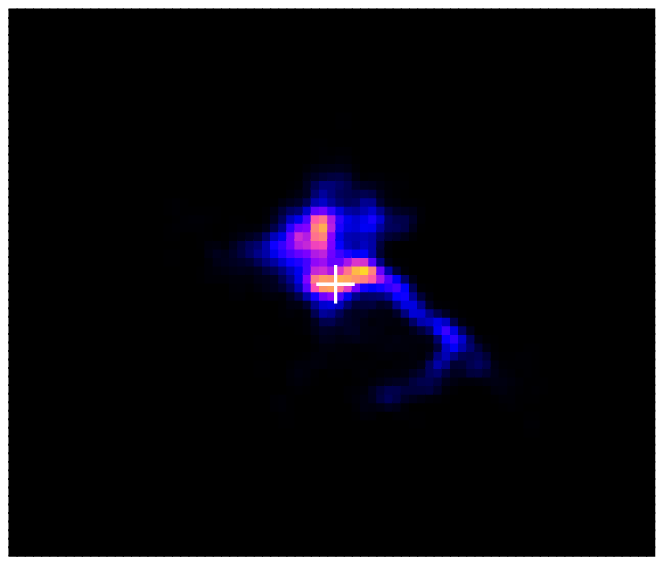}
\includegraphics[width=0.325\textwidth]{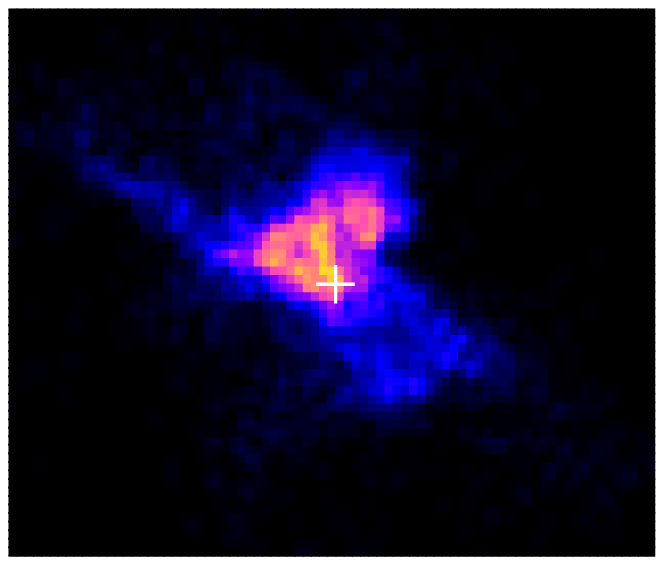}
\includegraphics[width=0.325\textwidth]{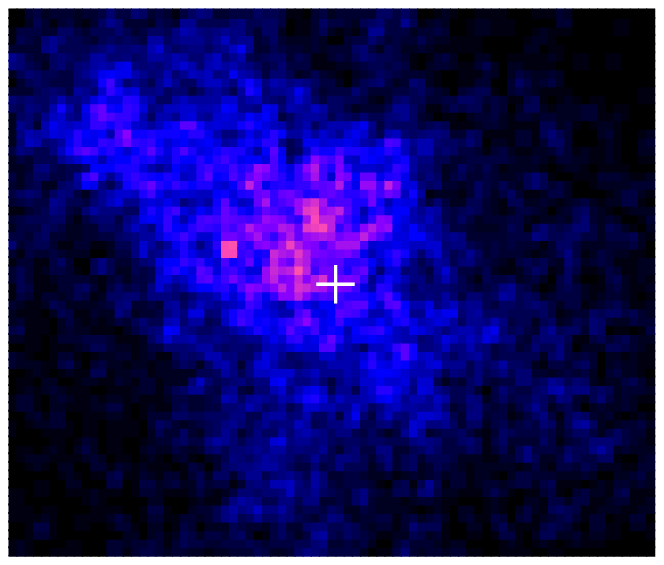}
\includegraphics[width=0.325\textwidth]{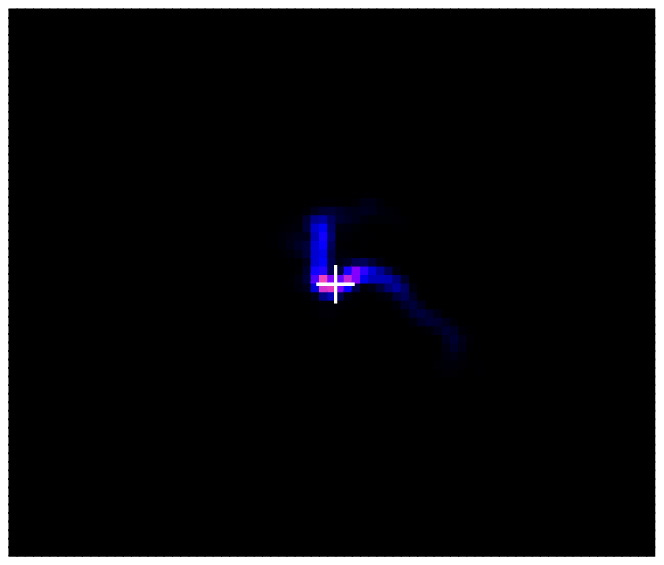}
\includegraphics[width=0.325\textwidth]{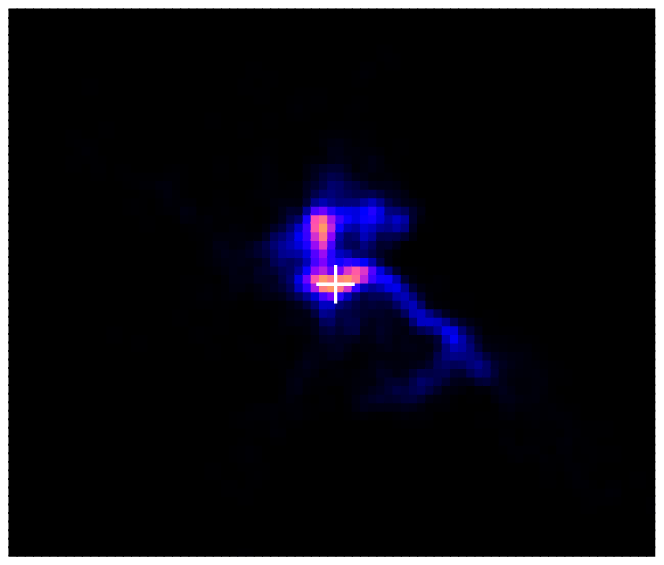}
\includegraphics[width=0.325\textwidth]{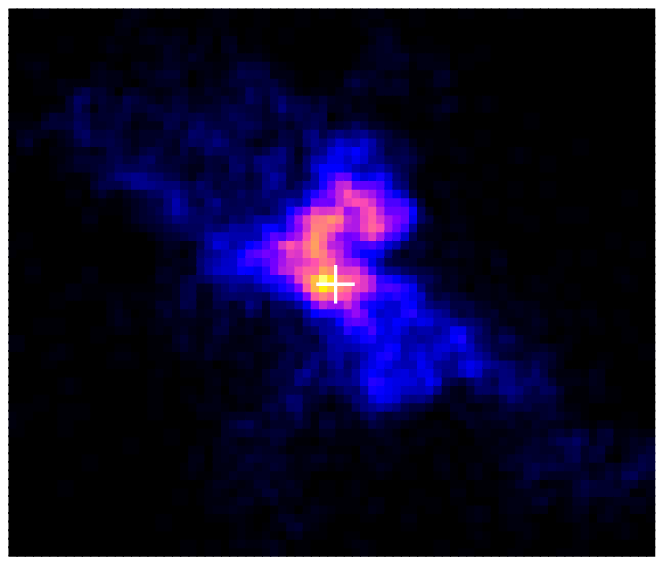}
\includegraphics[width=0.325\textwidth]{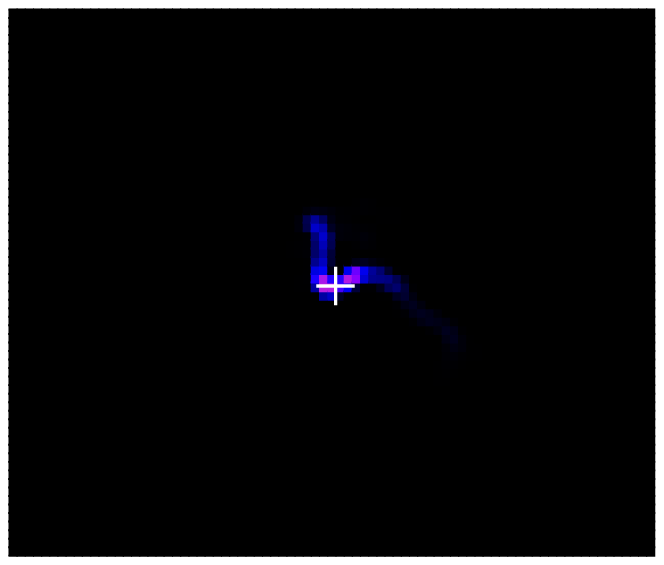}
\includegraphics[width=0.325\textwidth]{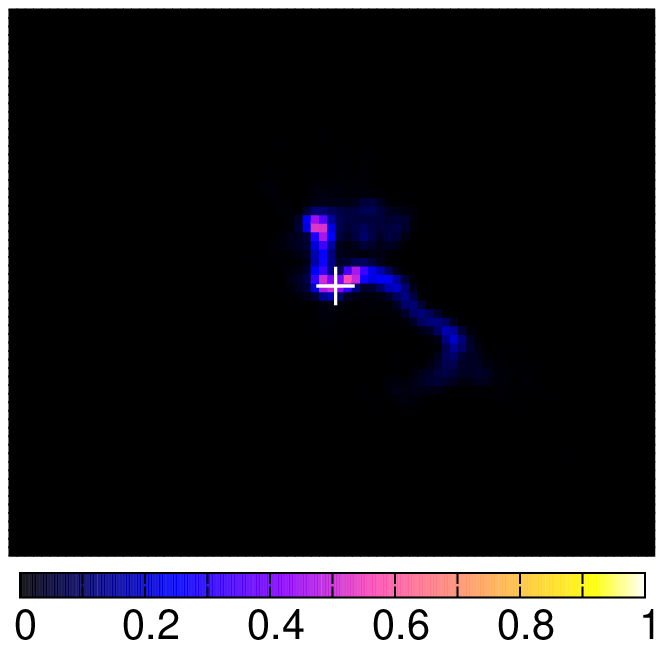}
\includegraphics[width=0.325\textwidth]{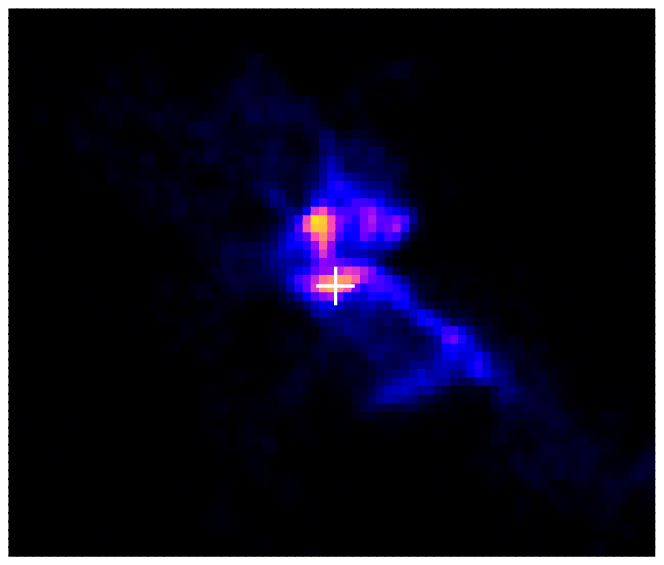}
\end{center}
\caption{Relative CR densities around a source, projected on 160\,pc$\times$160\,pc panels. CRs with $E/Z=1000$, 100 and 10\,TeV rigidities {\it (upper, middle and lower rows respectively)}, at times $t=0.5$, 2, 7\,kyr after emission {\it (left, middle and right columns respectively)}. White cross in the center for the position of the source.}
\label{ThreeE}
\end{figure}

We plot in Fig.~\ref{ThreeE} the projection on a plane of the CR distribution around a given source for different times and energies. In the upper row, $E/Z=1$\,PeV and $t=0.5$, 2, 7\,kyr. Such results are in line with those of Fig.~\ref{Eigenval} (left panel): Diffusion is initially strongly anisotropic ---if not filamentary, and then slowly tends towards the predictions of isotropic diffusion. Deviations, in particular from isotropy have also been found in simulations of energetic protons propagating in more or less anisotropic turbulence, see Ref.~\cite{Pommois2007}.

The two lower rows of Fig.~\ref{ThreeE} present CR distributions at lower rigidities ($E/Z=100$ and 10\,TeV). For decreasing $E/Z$, $t_{\ast}$ increases. From the similar shapes of distributions in panels on diagonals, one can see that the expected scaling $t_{\ast} \propto 1/D(E) \propto E^{-1/3}$ is approximately satisfied. We also find that $t_{\ast}$ approximately grows as $L_{\max}^{2}$. We estimate $t_{\ast} \sim 10\,{\rm kyr}\,(L_{\max}/150\,{\rm pc})^{\beta} ((E/Z)/{\rm PeV})^{-\gamma} (B_{\rm rms}/4\,\mu G)^{\gamma}$, where $\beta \simeq 2$ and $\gamma=0.25$--$0.5$. Secondary gamma-rays from CRs somehow map the CR distribution around the sources. For a uniform density of thermal protons in the surrounding ISM, the gamma-ray images would be similar to those of the CR column density, such as in Fig.~\ref{ThreeE}.

The above similarities between CR distributions at different energies, through the scaling $t_{\rm after \,\, escape} \propto 1/D(E)$, have been found when considering CRs as test particles. However, CR-driven instabilities may modify interstellar magnetic fields. CRs amplify and modify magnetic fields just ahead of supernova remnant shocks, see~\cite{Bell:2013kq} for a recent study. However, CR currents are significantly lower in the case studied here. Whether CRs would still have a non-negligible impact on fields within $\sim l_{\rm c}$ from their sources depends on several parameters, such as the amount of escaping CRs. If so, the anisotropy of the CR distribution would be lowered, and this should be expected to happen preferentially at lower energies where the CR current is larger. Our order-of-magnitude estimate below shows that the above test particle calculations should be sufficient for, at least, CRs with energies larger than $\sim$~a few tens of TeV, and therefore for photons with energies above a few TeV. Deviations would be larger in regions with strong cosmic ray currents $j$. Largest $j$ are mainly expected at $t<t_{\ast}$ when CRs are still close to their sources, and contained in well collimated flux tubes. For a source that has channeled $10^{50}$\,erg in CRs with $E=1\,{\rm GeV}-1\,{\rm PeV}$ and with an $E^{-2}$ spectrum, $10^{49}$\,erg would be present in each of the 10 bins in energy with logarithmic widths. Let us assume the extreme case where CRs are contained in a collimated tube of radius 3\,pc and length $2L_{\rm c}=2L_{\max}/5=60$\,pc, around the source. This yields a CR density $U \sim 100$\,eV/cm$^{3}$. CRs propagate inside with a speed roughly $\sim D/L_{\rm c}$ where $D=D_{0}E^{1/3}$ and $D_{0} \sim 10^{29}$\,cm$^{2}$/(s$\cdot$PeV$^{1/3}$)~\cite{Giacinti:2012ar}. The non-resonant hybrid (NRH) instability uncovered by Bell~\cite{Bell2004} dominates over the Alfv\'en instability when $Bjr_{\rm L}/\rho v_{\rm A}^{2}>1$, where $\rho \simeq 1\,m_{\rm p}\,$/cm$^{3}$ is the density of the ISM. For the above values, we are at the limit where it may play a role. The respective growth rates of these instabilities are~\cite{Bell2004} : $\gamma_{\rm NRH} = 0.5j\sqrt{\mu_0/\rho}$ and $\gamma_{\rm A} \approx 0.3j\sqrt{\mu_0/\rho}$.

Our parameters lead to a typical growth time of $\simeq 5 \gamma_{\rm NRH}^{-1} \approx 10\sqrt{\rho/\mu_0}\,E^{2/3}L_{\rm c}/UeD_0$. $5 \gamma_{\rm NRH}^{-1} \approx 3.1$ and 67\,kyr for CRs with $E=10$\,TeV and 1\,PeV, respectively. The bulk of CRs roughly spend a time $\sim L_{\rm c}^{2}/D$ in the collimated flux tube. Then, instabilities cannot grow sufficiently for CRs with roughly $E > \sqrt{\mu_0/\rho}\,UeL_{\rm c}/10 \approx 40$\,TeV. This energy may be further lowered if waves are damped sufficiently quickly. See for example~\cite{Bell1978,Damping,Blasi:2012yr} for sources of damping. If the impact of low energy CRs on magnetic fields is non-negligible, we expect the parallel diffusion coefficient along the filament to be suppressed, and CRs to diffuse more isotropically. $j$ and the growth rate $5 \gamma_{\rm NRH}^{-1}$ would become lower. Hence, some anisotropies in the CR distribution should be expected to remain at TeV energies even in such a case. Assuming a given template for CR escape from the source, one may check in the future if CR-driven instabilities have time to grow at low $E$ and have an impact on the surrounding fields, by 'comparing' gamma-ray images at low and high energies. At high energies, $\gamma$-ray observations will improve our knowledge of the structure of interstellar magnetic fields.

\section{Conclusions and perspectives}

We have discussed here two effects that are not predicted by the DA in its usual form.

First, we have proposed a natural explanation for the TeV--PeV CR anisotropies observed at small scales on the sky. The DA cannot predict them because it is not applicable on spatial scales below the CR mean free path. We have demonstrated in Sec.~\ref{CRASAS} that SSA must automatically appear due to the given local configuration of the magnetic field within $\approx l_{\rm mfp}$ from the observer, provided a dipole anisotropy exists. In the future, TeV--PeV CR anisotropies should become a convenient way to probe the structure of interstellar magnetic fields within a few tens of parsecs from Earth. The argument presented here holds for any field within $\approx l_{\rm mfp}$ from the observer. At $E \sim 1 - 10$\,TeV, the SSA may start to probe heliospheric fields, see also~\cite{Desiati:2011xg}. Let us mention that electric fields in the heliosphere may result in SSA too, see~\cite{Drury:2013uka}. In the future, a thorough analysis can determine the relative contributions of both effects.

Second, we have shown in Sec.~\ref{ACRD} that diffusion of TeV--PeV CRs in the ISM should be expected to be non-negligibly anisotropic on scales smaller than $L_{\max}\sim 100$\,pc. Therefore, CR distributions around recent sources should look anisotropic and irregular. This has important implications for $\gamma$-ray astronomy, and some first observations may hint at our findings, see~\cite{Giacinti:2013wbp}. Detailed comparisons at high and low photon energies of the extended $\gamma$-ray emissions around CR sources may give insights into the potential impact at low energies of CR-driven instabilities on the surrounding ISM.

In general, deviations from standard diffusion are expected to have an impact in several other situations. For example, Refs.~\cite{TSA} consider, in particular, their implications for particles accelerated at the solar wind termination shock.

\section*{Acknowledgments}

The author acknowledges funding from the European Research Council under the European Community's Seventh Framework Programme (FP7/$2007-2013$)/ERC grant agreement no. 247039.

\end{document}